\documentclass[twocolumn,trackchanges]{aastex6}
\usepackage{gensymb}
\usepackage{natbib}
\usepackage{lineno}
\begin{document}
\turnoffedit
\submitted{}
\title{Endogenic and Exogenic Contributions to Visible-wavelength Spectra of Europa's Trailing Hemisphere}

\author{Samantha K. Trumbo, Michael E. Brown}
\affil{Division of Geological and Planetary Sciences, California Institute of Technology, Pasadena, CA 91125, USA}
\author{Kevin P. Hand}
\affil{Jet Propulsion Laboratory, California Institute of Technology, Pasadena, CA 91109, USA}

\begin{abstract}
The composition of Europa's trailing hemisphere reflects the combined influences of endogenous geologic resurfacing and exogenous sulfur radiolysis. Using spatially resolved visible-wavelength spectra of Europa obtained with the Hubble Space Telescope, we map multiple spectral features across the trailing hemisphere and compare their geographies with the distributions of large-scale geology, magnetospheric bombardment, and surface color. Based on such comparisons, we interpret some aspects of our spectra as indicative of purely exogenous sulfur radiolysis products and other aspects as indicative of radiolysis products formed from a mixture of endogenous material and magnetospheric sulfur. The spatial distributions of two of the absorptions seen in our spectra---a widespread downturn toward the near-UV and a distinct feature at 530 nm---appear consistent with sulfur allotropes previously suggested from ground-based spectrophotometry. However, the geographies of two additional features---an absorption feature at 360 nm and the spectral slope at red wavelengths---are more consistent with endogenous material that has been altered by sulfur radiolysis. We suggest irradiated sulfate salts as potential candidates for this material, but we are unable to identify particular species with the available data. 
\end{abstract}


\section{Introduction}\label{sec:intro}
\begin{figure*}
\figurenum{1}
\plotone{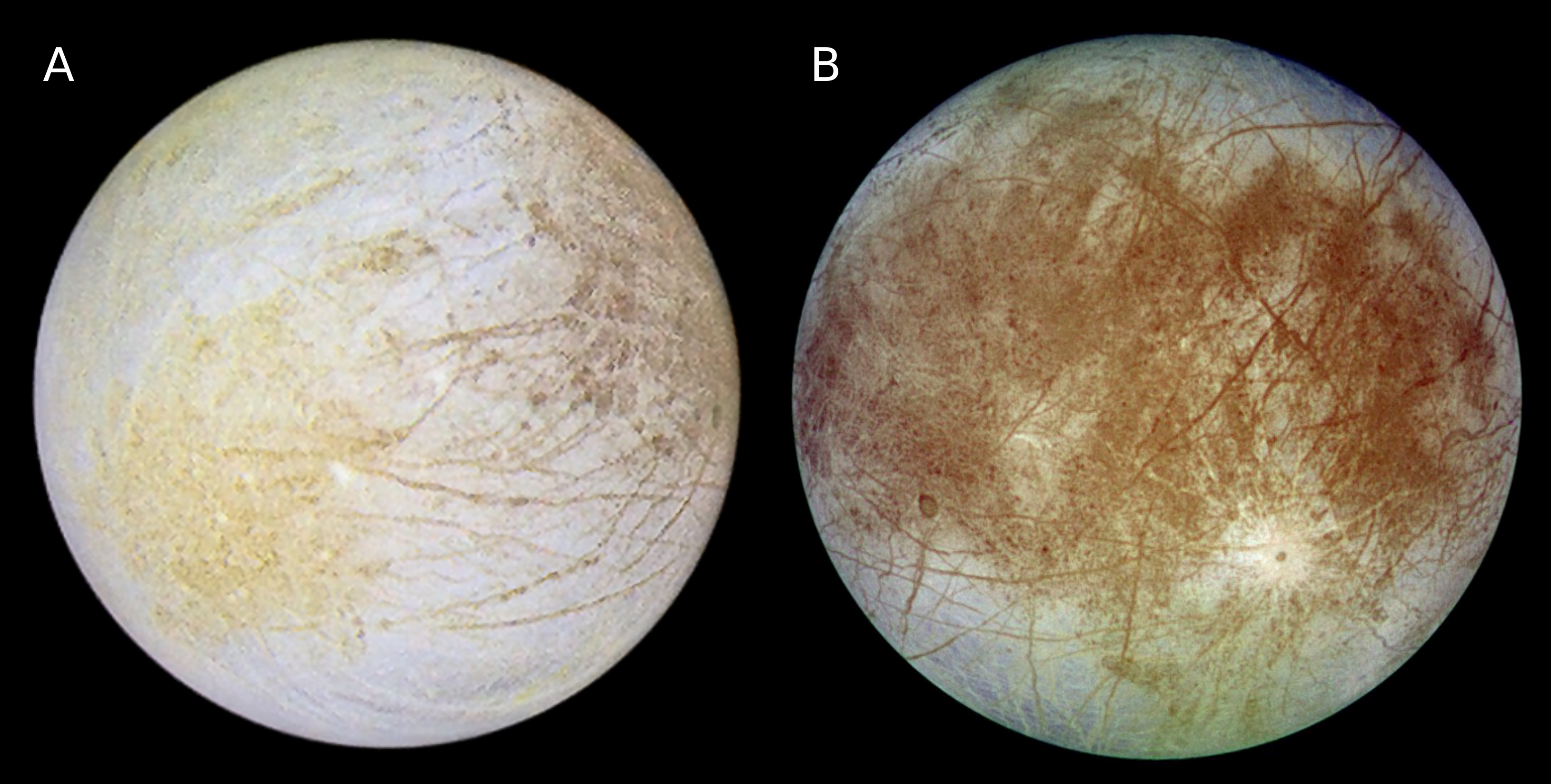}	
\caption{\textit{Galileo} SSI color images of approximately the leading (A) and approximately the trailing (B) hemispheres (PIA01295 and PIA00502 in the NASA JPL Photojournal). The actual central longitudes of the images are closer to 45$\degree$W and 295$\degree$W, respectively. These approximate true-color images were created using the \textit{Galileo} violet, green, and near-infrared (986 nm) filters. Both images show a clear association of color with geologic features, though the geology of the trailing hemisphere appears significantly redder than its more yellow leading-hemisphere counterparts. Individual lineae that traverse from the trailing to the leading hemisphere change color from red to yellow as they leave the sulfur-implantation experienced on the trailing hemisphere. The surface color's simultaneous correlation with geology and dichotomy between the hemispheres suggest that the color may indicate endogenous material on the leading hemisphere and endogenous material altered by sulfur radiolysis on the trailing hemisphere. The large yellow patch in the lower left of the leading-hemisphere image is the large-scale chaos region Tara Regio, where HST spectra detect irradiated NaCl \citep{TrumboEtAl2019}. Image credits: NASA/JPL/University of Arizona. \label{fig:gal_images}}
\end{figure*}

Images of Europa from the \textit{Voyager} and \textit{Galileo} spacecrafts show striking color variations across the surface that exhibit marked hemispherical differences and correlations with surface geology \citep[e.g.][]{Johnson1983,McEwen1986,Buratti1988,Clark1998,Fanale1999}. These visible patterns likely reflect the combined influences of endogenous and exogenous sources on the underlying surface composition. A unique association of color with geologic features, such as lineae and heavily disrupted ``chaos" terrain (Figure \ref{fig:gal_images}), pervades the entire surface and hints at the possibility that compositional fingerprints of the internal ocean may persist within recent geology. However, a distinct color contrast between the leading and trailing hemispheres, in which the geologic features of the trailing hemisphere are significantly darker and redder than their leading-hemisphere counterparts (Figure \ref{fig:gal_images}), appears to reflect the constant exogenous alteration of the trailing-hemisphere surface chemistry via sulfur radiolysis \citep{McEwen1986, Nelson1986, Johnson1988, Carlson2009}. Sulfur plasma ions from the volcanos of Io co-rotate with Jupiter's magnetic field and continuously deposit onto the trailing hemisphere \citep{PospJohnson1989, Paranicas2009}, where bombardment by energetic magnetospheric electrons, protons, and ions \citep{Paranicas2001,Paranicas2002,Paranicas2009} drives a chemically active radiolytic sulfur cycle that affects the underlying composition \citep{Carlson2002, Carlson2005}. Indeed, continuous lineae that traverse from the trailing to the leading hemisphere appear to change color, becoming less red as they become sheltered from the impinging sulfur plasma (Figure \ref{fig:gal_images}). Such exogenic processing complicates the interpretation of surface components as oceanic signatures, even within geologically young terrain. Disentangling potential endogenous species from radiolytic products is thus critical to understanding the surface composition of Europa and thereby constraining the chemistry of the ocean below.

The imagery implies that visible wavelengths contain compositional information, which may help distinguish endogenic from exogenic influences. Indeed, multiple studies have utilized broadband photometry and spectral ratios from these images to reveal patterns in visible reflectance associated with plasma bombardment and geologic units \citep{McEwen1986, Nelson1986, Johnson1988, Buratti1988,Clark1998,Fanale1999}. However, as neither \textit{Voyager} nor \textit{Galileo} carried a visible-wavelength spectrometer, such studies lacked detailed spectral information that could provide further insight into the compositional differences responsible for the patterns observed.

Until recently, visible spectroscopy of the surface has been limited to disk-integrated observations obtained from the ground \citep[e.g][]{Johnson1970, Johnson1970_thesis, McFadden1980, Spencer1995, Carlson2009}. These spectra echo the leading/trailing albedo and color contrasts seen in imagery and reveal some notable spectral features, including possible absorptions near 360 and 530 nm on the trailing hemisphere and a broad, global downturn toward the near UV (with a band edge at $\sim$500 nm) that is stronger on the trailing hemisphere. However, despite the fact that Europa's surface color shows a clear association with geology, suggesting endogenous influences at visible wavelengths, the features visible in the ground-based spectra have most often been attributed entirely to sulfur allotropes and SO$_2$ \citep{Spencer1995, Carlson2009}. Though it was suggested that some sulfur could be endogenic, these species are also anticipated products of the exogenic sulfur implantation \citep{Steudel1986, Carlson2002, Carlson2009}, which is indiscriminate of underlying geology.

More recent thinking, however, has considered the possible visible-wavelength contributions of salts related to the internal ocean, which would more plausibly follow disrupted terrain and can become visibly colored due to the formation of radiation-induced defects known as ``color centers" \citep{HandCarlson2015, Poston2017, Hibbitts2019}. Distinguishing between the potential spectral signatures of salts and sulfur products may be possible with spatially resolved spectroscopy, which can isolate large-scale geologic regions. Indeed, spatially resolved visible-wavelength spectra taken with the Hubble Space Telescope (HST) have already revealed what appears to be a color-center absorption of irradiated sodium chloride (NaCl) at 450 nm on the leading hemisphere, challenging the idea that Europa's surface color and visible spectrum solely reflect sulfur species \citep{TrumboEtAl2019}. The NaCl feature appears exclusively on the leading hemisphere, separate from the trailing-hemisphere sulfur radiolysis, and correlates with surface geology and color, corresponding particularly to Tara Regio, a large, visibly yellow region of chaos terrain (Figure \ref{fig:gal_images}). NaCl may explain some of the visible patterns on the leading hemisphere, but the species responsible for those on the trailing hemisphere remain uncertain. Here, we use the same HST visible-wavelength dataset to investigate the composition of the trailing hemisphere. We map visible spectral features across the surface and compare their geographic distributions with surface geology, surface color, and particle bombardment patterns in an attempt to distinguish between endogenic and exogenic origins. 

\section{Observations and Data Reduction}\label{sec:observations}
We observed Europa with the Space Telescope Imaging Spectrograph (STIS) across four HST visits in 2017. The corresponding dates, times, and geometries are listed in Table \ref{table:obs}. During each visit, we repeatedly stepped the 52$^{\prime\prime}$ x 0.1$^{\prime\prime}$ slit in 0.06$^{\prime\prime}$ increments across the full disk of Europa, resulting in overlapping aperture positions. We executed this slit-scan pattern twice per visit---once each in the G430L and G750L first-order spectroscopy modes (R $\sim$500) to achieve full 300--1000 nm wavelength coverage. At each slit position, we integrated for either 9 (G750L) or 10 seconds (G430L). Flux- and wavelength-calibrated data were then provided by HST after standard reduction with the STIS calibration pipeline (calstis). Using the same pipeline, but including the calstis defringing procedures, we reprocessed the G750L data to remove substantial fringes from the longest wavelengths. We extracted single spectra by taking individual rows from the two-dimensional spectral images, corresponding to the 0.05$^{\prime\prime}$ pixel-scale ($\sim$150-km diffraction-limited resolution at 450 nm). We then divided each spectrum by the ASTM E-490 solar reference spectrum \citep{SolarRef2000} to convert to reflectance. 

\begin{table}
\begin{center}
\caption{Table of Observations\label{table:obs}}
\begin{tabular}{cccc}
\hline\\[-4mm] \hline
Date&Time&Central&Central\\
(UT)&(Start/End)&Lon.&Lat.\\ \hline
2017 May 13 & 00:20/01:41 &  224 & -3.06\\
2017 Jun 29 & 08:19/08:55 &  25 & -2.91\\
2017 Aug 1 & 04:43/05:20 &  133 & -2.91\\
2017 Aug 6 & 12:27/13:47 &  314 & -2.92\\ \hline
\end{tabular}
\end{center}
\end{table}

The G750L data ($\sim$550--1000 nm) seemed to contain multiple artifacts, some of which may have been residuals of the defringing process similar to those seen in STIS spectra of Mars \citep{Bell2007}. In addition, significant slit losses and the broad point spread function of STIS distorted the continuum spectral shape in the G750L setting. To correct for these effects, we fit a spline curve to a high-quality ground-based spectrum of the leading hemisphere \citep{Spencer1995} and extended the fit as a constant beyond the extent of the ground-based spectrum ($\sim$775 nm), which is approximately consistent with spectrophotometric measurements at these wavelengths \citep{McFadden1980}. We then multiplied our spectra by the ratio of this curve to a corresponding disk-integrated spectrum constructed from our G750L data. This approach simultaneously divided out global artifacts from the G750L spectra and corrected the continuum shape for slit losses, while preserving relative differences between individual spectra. Finally, to produce continuous 300--1000 nm spectra of the entire surface, we combined the G430L and G750L settings, scaling as appropriate to correct minor flux offsets and smoothing the G430L data to match the G750L signal-to-noise. We calculated the corresponding latitude/longitude coordinates of each extracted pixel using the known phase and angular size of Europa (as obtained from JPL Horizons) and the aperture geometry information included in the HST FITS headers.

\section{Spectral Maps}\label{sec:mapping}
\begin{figure}
\figurenum{2}
\plotone{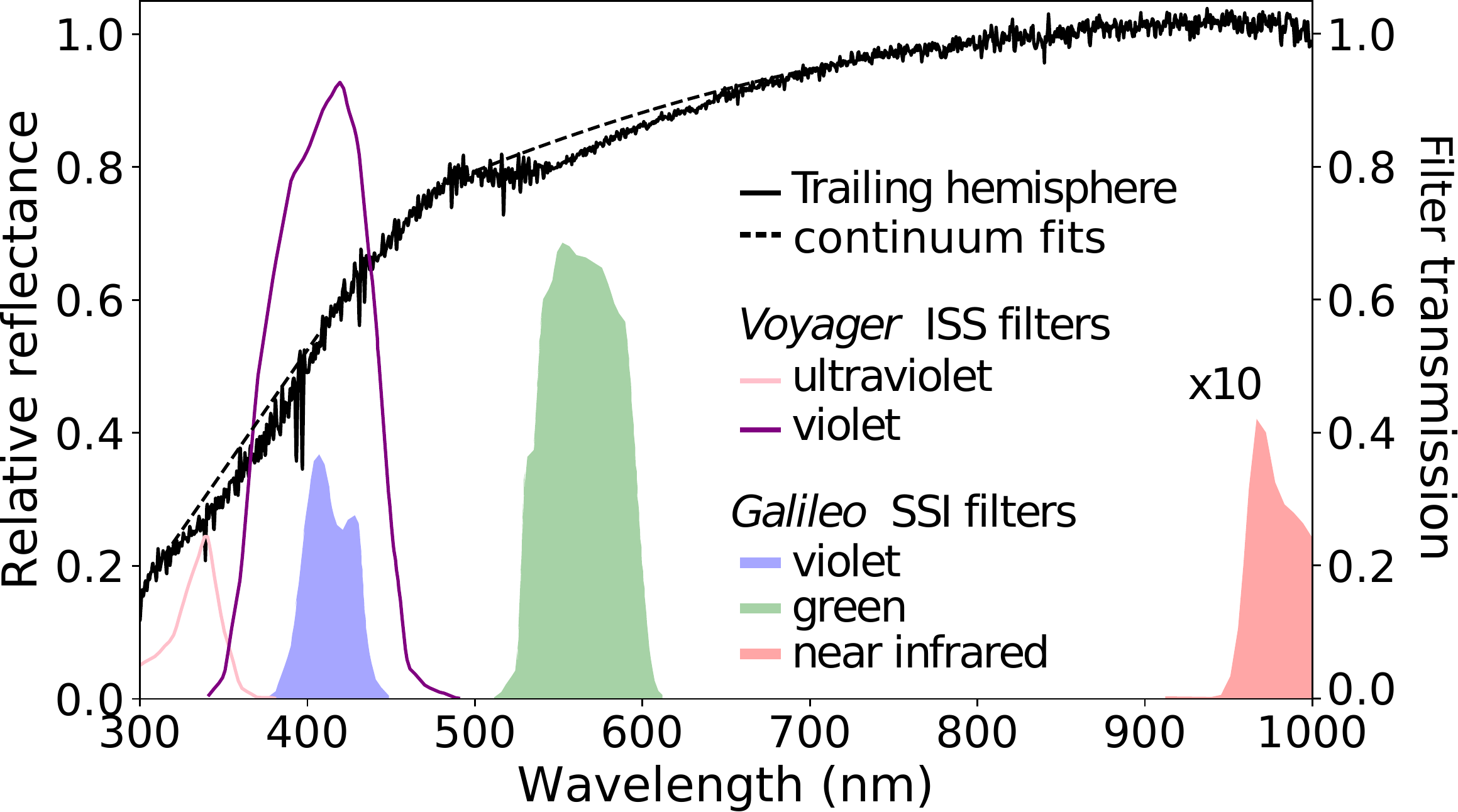}
\caption{Representative spectrum from our HST data of the trailing hemisphere of Europa compared to the \textit{Voyager} and \textit{Galileo} imaging filters. The spectrum is an average from Eastern Annwn Regio and features a strong near-UV downturn with a band edge near 500 nm, as well as two discrete features near 360 and 530 nm. Black dashed lines indicate representative continuum fits akin to those used to map the strength of each feature in our individual spectra. We include the \textit{Voyager} ultraviolet and violet filter responses underneath the spectrum, as well as the \textit{Galileo} filters used to create the images in Figure \ref{fig:gal_images}. The \textit{Galileo} near-infrared (986 nm) filter response is multiplied by 10 for clarity.\label{fig:trailing}}
\end{figure}

Our spectra of the trailing hemisphere (Figure \ref{fig:trailing}) show the same strong downturn toward the near UV (with a band edge around 500 nm) that was seen in prior ground-based spectrophotometry, and better spectrally resolve the discrete features near 360 and 530 nm that were more tentatively detected \citep{Johnson1970, Johnson1970_thesis, McFadden1980, Carlson2009}. Previously, it was suggested that an assortment of sulfur allotropes could explain all three features, with the 360 and 530 nm absorptions tentatively identified as S$_\mu$ (polymeric sulfur) and S$_4$ (tetrasulfur), respectively, and the broad near-UV downturn most often associated with $\alpha$-S$_8$ (orthorhombic cyclo-octal sulfur) \citep{Spencer1995,Carlson2009}. In one respect, invoking sulfur allotropes to explain the visible spectrum of the trailing hemisphere makes sense due to the sulfur implantation and radiolysis known to be occurring there. However, the imagery clearly implies that some aspects of the visible spectrum must be related to geology, which one would not necessarily expect of radiolysis products composed of pure sulfur. In order to investigate which aspects of our spectra may be endogenous in origin and which can be attributed to exogenous sulfur chemistry, we map the strength of the aforementioned features across the surface and look for correlations with surface color, geology, and radiation bombardment patterns.

To independently measure the strength of the discrete 360 nm absorption and of the larger-scale near-UV downturn on which it is superimposed, we normalize each spectrum to the median reflectance of the 415--425 nm region and fit a linear continuum from 307.5 to 425 nm, excluding the portion corresponding to the discrete absorption ($\sim$315--415 nm). We assess each fit by eye and, if necessary, make small changes to these bounds. We take the slope of the fitted continuum as a measure of the magnitude of the near-UV downturn. We then divide out the calculated continuum from each spectrum and integrate the residual absorption to obtain the \edit1{band area} of the 360 nm feature. We take a similar approach to measure the \edit1{band area} of the 530 nm feature, instead using a second-order polynomial continuum between $\sim$480 and 770 nm, excluding the wavelengths of the apparent absorption ($\sim$500--700 nm) and making adjustments when necessary to achieve a satisfactory continuum fit. Representative continua are included in Figure \ref{fig:trailing}. Finally, we map our measures of all three absorptions across the surface using the geographic coordinates as obtained in Section \ref{sec:observations}. We exclude data near the limb of Europa, as the spectra are of poorer quality, making accurate quantification of spectral features difficult.

Figure \ref{fig:maps} shows the results of this mapping compared to the \textit{Voyager} UV/VI ratio map \citep{McEwen1986, Carlson2005, Carlson2009}, which was constructed from images taken in the \textit{Voyager} ultraviolet (UV) and violet (VI) filters. The \textit{Voyager} UV/VI map (Figure \ref{fig:maps}A) has long been interpreted to primarily reflect the effects of exogenous sulfur implantation on the trailing hemisphere, as the large-scale pattern of UV dark material forms an elliptic pattern centered around the trailing point (0$\degree$N, 270$\degree$W) that largely coincides with the expected patterns of both Iogenic sulfur and electron bombardment \citep{PospJohnson1989,Paranicas2001, Paranicas2009}. Indeed, like the expected sulfur flux, the \textit{Voyager} UV/VI ratio varies roughly as the cosine of the angle from the trailing point, though the relationship is not perfectly linear \citep{Nelson1986, McEwen1986}.  However, as \citet{McEwen1986} noted, the UV/VI map also features smaller-scale patterns that appear to be endogenic in origin and that precisely associate with geology. In particular, the large-scale chaos regions Dyfed Regio ($\sim$250$\degree$W) and Eastern Annwn Regio ($\sim$294$\degree$W) and the intervening smaller-scale chaos regions appear especially dark in the UV/VI map, but discrete features south of Pwyll Crater (25$\degree$S, 271$\degree$W) also appear distinct from the background elliptic pattern. In fact, in comparing the \textit{Voyager} ultraviolet and violet filter responses \citep{Danielson1981} to a representative trailing-hemisphere spectrum (Figure \ref{fig:trailing}), we see that the UV/VI ratio simultaneously measures two different things---the large near-UV downturn and the discrete 360 nm feature. Our analysis attempts to separate the two. 

\subsection{Near-UV downturn} \label{subsec:nuv}
We find that mapping the slope across the 315--415 nm region (our proxy for the near-UV downturn) reproduces the large-scale, apparently exogenic pattern of the UV/VI map. With the exception of a few spuriously strong slopes near the northern limbs of each observation, which we believe are pixel-dependent artifacts, the slopes on the trailing hemisphere follow a largely uniform and symmetric elliptic distribution centered on the trailing point and tapering toward the sub- and anti-Jovian points (Figure \ref{fig:maps}B). Again, this pattern is largely consistent with the expected geographies of sulfur implantation and electron bombardment on the trailing hemisphere \citep{PospJohnson1989,Paranicas2001, Paranicas2009}, suggesting an exogenic origin for the near-UV downturn. It is worth noting, however, that this slope is not a perfect measure of the near-UV downturn everywhere across the surface, as it is disrupted by the 450 nm NaCl absorption on the leading hemisphere \citep{TrumboEtAl2019}. Indeed, the NaCl feature, which falls partly within the \textit{Voyager} violet filter and is strongest in the large-scale chaos region Tara Regio (10$\degree$S, 75$\degree$W), explains much of the red ``UV-bright" material in the \textit{Voyager} UV/VI map and results in a depressed slope by our measure. In reality, this region also exhibits an overall drop in reflectance toward the near-UV that is comparable to that of the immediately surrounding terrain. In fact, though the near-UV downturn is strongest on the trailing hemisphere, all of our spectra exhibit a downturn toward the near UV, and the presence of an absorption edge at $\sim$500 nm appears to be a truly global characteristic that is independent of terrain type. Thus, while the strong near-UV downturn on the trailing hemisphere certainly appears to result from the exogenous sulfur chemistry, potentially reflecting the previously suggested sulfur allotrope $\alpha$-S$_8$ or some combination of sulfur allotropes that absorb strongly in the UV, alternative explanations may be worth considering for the \edit1{weaker} near-UV downturn observed elsewhere. Indeed, the near ubiquitous presence of an absorption edge near 500 nm on the other icy Galilean satellites \citep{Spencer1995} as well as on the icy Saturnian satellites \citep{Hendrix2018} supports this idea. Radiation-processed organics are invoked to explain the near-UV downturn on the Saturnian satellites \citep{Hendrix2018}. However, limited laboratory data have suggested that radiation-damaged water ice could exhibit a similar near-UV downturn \citep{Sack1991}, which perhaps presents \edit1{an alternative explanation for the leading hemisphere and icy regions of Europa}, as there is currently no evidence for widespread organics at other wavelengths.

\subsection{360 nm feature} \label{subsec:360}
Our map of the discrete 360 nm band (Figure \ref{fig:maps}C) reveals a more irregular and spatially localized pattern that is strongest near the trailing point, but that does not fill the entire elliptic pattern of exogenous alteration. Instead, the geographic distribution of the 360 nm feature appears to correspond to the same geology as the endogenic patterns visible in the \textit{Voyager} UV/VI map, but simply mapped at the coarser spatial resolution of our HST data. Like the lowest \textit{Voyager} UV/VI ratios, the strongest 360 nm absorptions appear associated with Dyfed Regio, Eastern Annwn Regio, and the intervening smaller-scale chaos terrain, with more moderate strengths south of Pwyll Crater. In fact, as the UV/VI ratio is necessarily decreased by the presence of the 360 nm feature, we can say with some certainty that our map of the 360 nm band strength reflects the same geologic regions. Indeed, applying the HST point spread function and pixel scale to a starting distribution corresponding to the lowest ratios in the Voyager map produces a pattern very similar to the geography of the 360 nm feature that we observe. 

The association with geologically young chaos terrain implies that the 360 nm feature reflects endogenous influences on the surface composition. However, its confinement to the sulfur-bombarded trailing hemisphere simultaneously suggests that it is related to the exogenous sulfur radiolysis occurring there. Indeed, the fact that the 360 nm absorption is not equally strong within all trailing-hemisphere chaos terrain, but is instead concentrated within that closest to the trailing point, suggests that it may depend heavily on the impinging sulfur flux. All together, this geography is suggestive of an endogenous material that has been compositionally altered by sulfur radiolysis. Previously, the 360 nm absorption was tentatively attributed to the sulfur allotrope S$_\mu$ \citep{Carlson2009}. However, as S$_\mu$ can likely result solely from the radiolysis of implanted Iogenic sulfur \citep{Steudel1986, Carlson2009}, requiring no endogenous input, there is no obvious reason to expect a correlation with chaos terrain. Thus, while it is conceivable that there may be unknown effects acting to concentrate or enhance the stability of S$_\mu$ within chaos regions, it is worth re-evaluating the cause of the 360 nm feature and considering species that are not pure sulfur, but that instead form radiolytically from a mixture of Iogenic sulfur and endogenic materials.

\subsection{530 nm feature} \label{subsec:530}
The 530 nm absorption proved more difficult to quantify, as it falls at the junction between the G430L and G750L settings and very near the $\sim$500 nm band edge of the near-UV downturn. Thus, the measurement of this feature was somewhat sensitive to slight slope and flux mismatches between settings, particularly at the limbs, as well as to changes in the near-UV absorption edge. As a result, our map of the 530 nm absorption is less certain, though mapping with different polynomial continua and fitting parameters consistently produces qualitatively similar geographies. We estimate the pixel-by-pixel uncertainty to be less than 1.5 nm \edit1{of band area} on average.

\begin{figure}
\figurenum{3}
\plotone{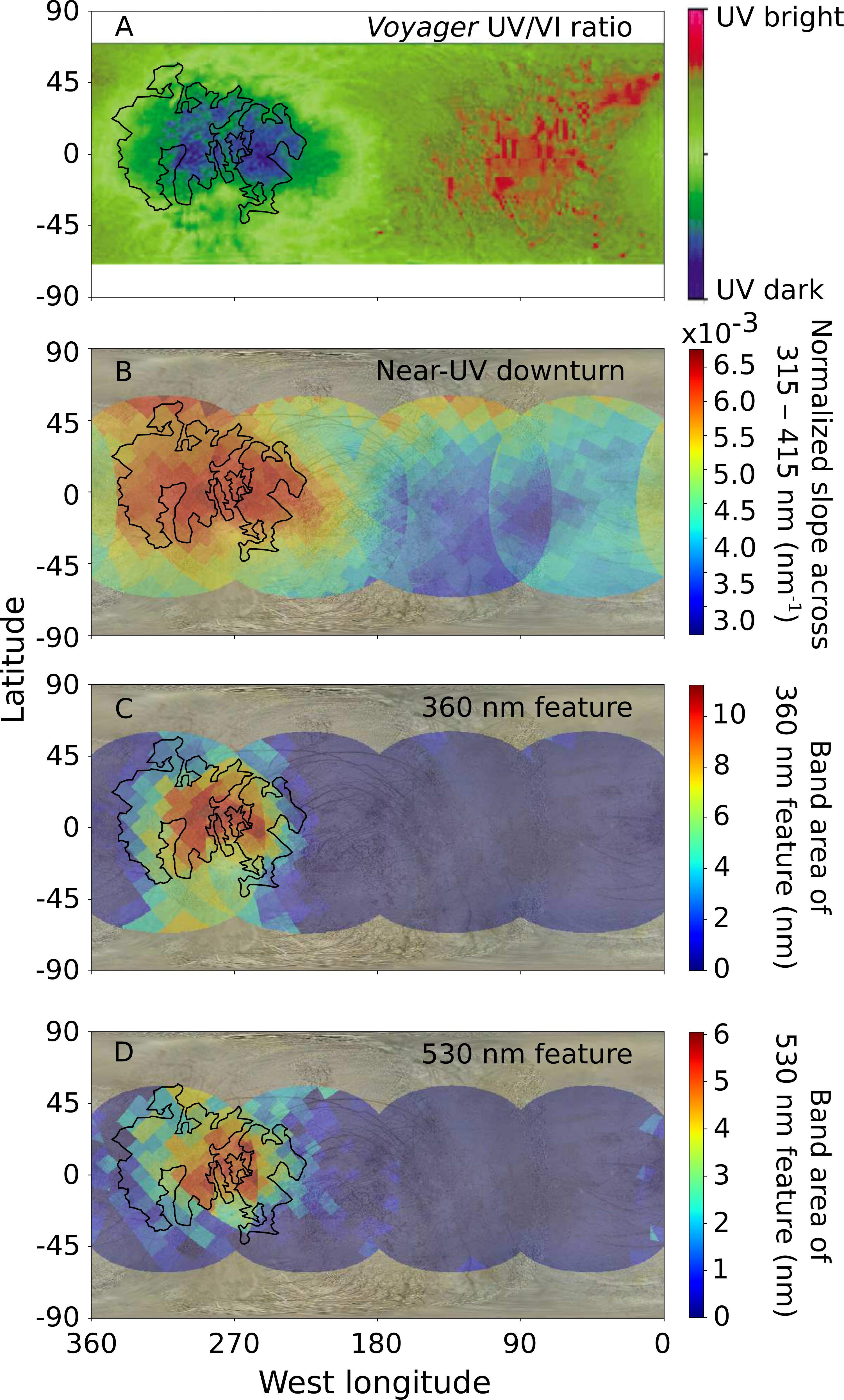}
\caption{(A) \textit{Voyager} UV/VI ratio map adapted from \citet{Carlson2005}, but originally produced by \citet{McEwen1986}. The large-scale elliptic pattern of UV-dark material on the trailing hemisphere likely reflects the exogenous sulfur chemistry occurring there. However, the UV/VI ratio also displays smaller-scale patterns associated with the large-scale chaos regions Dyfed Regio and Eastern Annwn Regio, the smaller-scale chaos terrain between them, and some apparent geology south of Pwyll Crater. Black outlines indicate the chaos terrain mentioned and are adapted from \citet{Doggett2009}. The responses of the \textit{Voyager} ultraviolet and violet filters used to create this map are included in Figure \ref{fig:trailing}. (B) Map of the slope from 307.5 to 425 nm (our proxy for the near-UV downturn) in our HST spectra, which reproduces the large-scale, exogenic pattern of the UV/VI map. This distribution suggests that the near-UV downturn reflects exogenous influences. (C) Map of the 360 nm band strength in our HST spectra, which resembles the smaller-scale, apparently endogenic portions of the \textit{Voyager} UV/VI map. This geography is suggestive of a combination of endogenic and exogenic influences. (D) Map of the 530 nm band strength in our HST spectra, which may be consistent with either an association with geology near the trailing point or with a simple dependence on the highest sulfur fluxes. \label{fig:maps}}
\end{figure}

The distribution we obtain (Figure \ref{fig:maps}D) is similar to that of the 360 nm feature in that it also displays the strongest absorptions near the trailing point and does not fill the entire exogenic alteration pattern. However, without a corresponding high-spatial-resolution imaging map sensitive to the 530 nm absorption, it is difficult to evaluate any potential correlation with the chaos terrain containing the 360 nm feature. Indeed, while such a correlation seems plausible from our map, the observed distribution of the 530 nm feature is also largely consistent with a simple concentration nearest the trailing point, which receives the highest sulfur flux. Thus, though it is possible that the 530 nm absorption also results from radiolytically altered endogenous material, it's previous identification as S$_4$ is equally consistent with our data.

\subsection{Correlations with visible color} \label{subsec:red_slope}
Though the strong near-UV downturn is widespread on the trailing hemisphere and at least the 360 nm feature correlates with some trailing hemisphere chaos terrain, none of the spectral features we have investigated thus far consistently correspond to the red color that appears common to all geology across the trailing hemisphere (Figure \ref{fig:gal_images}). The near-UV elliptic pattern overprints much of the underlying geologic features, but is significantly more uniform and more symmetric about the trailing point than is the visibly red large-scale geology, which is asymmetric and offset west from the apex. In contrast, the 360 nm feature does associate specifically with some of this geology, particularly Dyfed Regio and the eastern portion of Annwn Regio nearest the trailing point, but it is much weaker within the western portions of Annwn Regio, which are similarly red in color to their eastern counterparts. The 530 nm absorption is equally constrained to the most central portions of the trailing hemisphere. Thus, while all three features necessarily influence the colors visible in the \textit{Voyager} and \textit{Galileo} imagery, none appear to be an underlying commonality specifically associated with the widespread red material. 

\begin{figure}
\figurenum{4}
\plotone{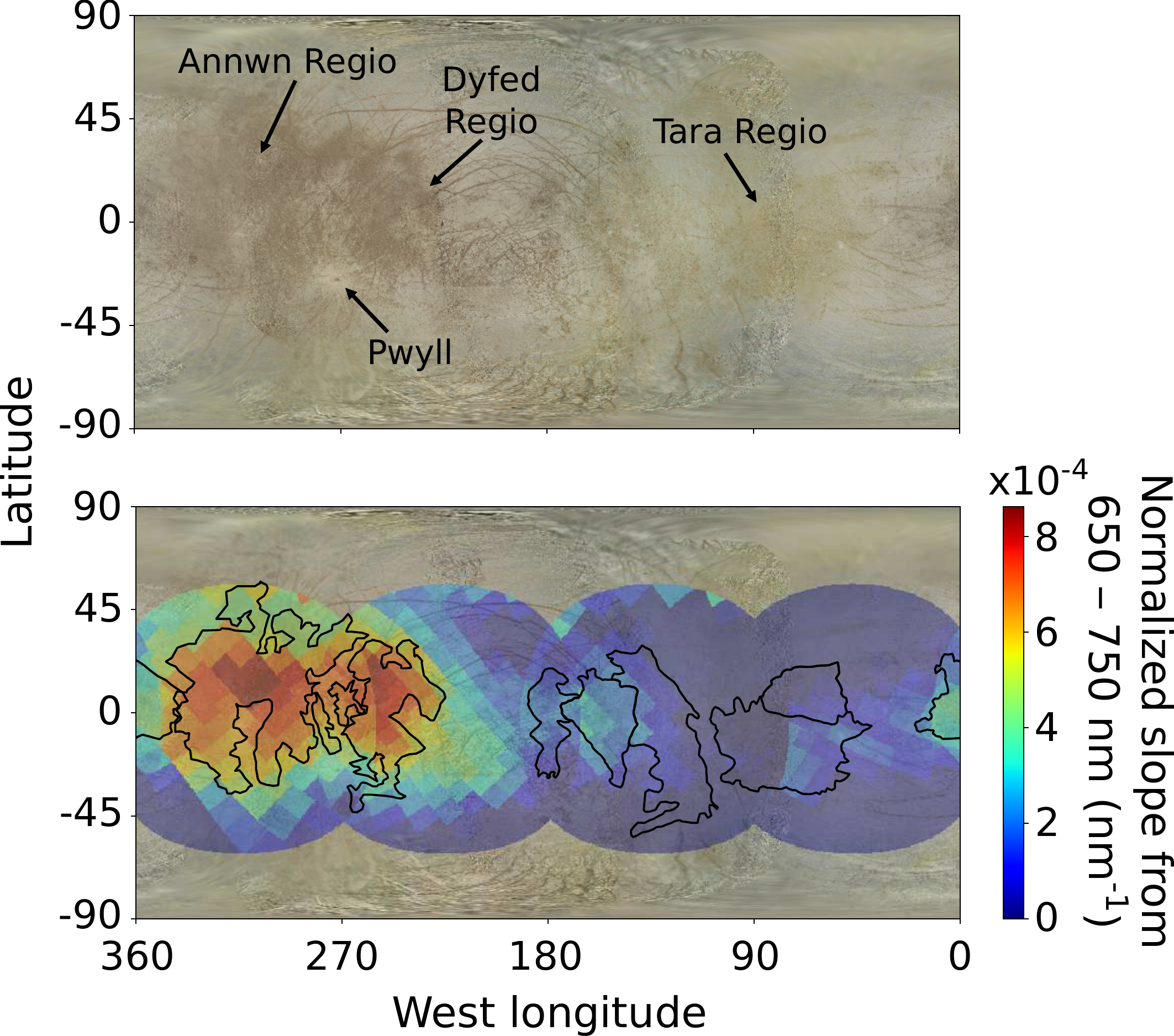}
\caption{Map of the spectral slope from 650 to 750 nm compared to an approximate true-color mosaic of Europa's surface (image credit: NASA / JPL / Bj\"{o}rn J\'{o}nsson). This slope acts as a measure of the broad absorption feature visible across the red wavelengths and corresponds well to the reddish material visible in the imagery. Our map of this slope highlights all of the large-scale trailing-hemisphere chaos terrain and even the less-red chaos regions near the sub- and anti-Jovian points to a lesser extent (black chaos outlines are adapted from \citet{Doggett2009}). As the broad absorption across the red wavelengths appears common to all of the large-scale geology experiencing sulfur radiolysis, it likely reflects species formed via the radiolysis of a mixture of endogenic material and implanted Iogenic sulfur. \label{fig:maps_color}}
\end{figure}

Instead, the aspect of our spectra that we find corresponds best geographically to the red material in the imagery is the slope in the 700 nm region. This slope appears to result from a broad absorption that extends through the red wavelengths before interfering with the 530 nm feature. As a proxy for its strength, we normalize our spectra to the median reflectance between 745 and 750 nm, linearly fit the data between 650 and 750 nm, and then map the resulting slopes across the surface. Figure \ref{fig:maps_color} shows the result of this mapping compared to an approximate true-color mosaic of Europa demonstrating the extent and locations of reddish material on the surface (image credit: NASA / JPL / Bj\"{o}rn J\'{o}nsson). Unlike any of the spectral maps discussed above, our map of this absorption seems uniquely correlated with all of the visibly red large-scale chaos terrain on the trailing hemisphere, highlighting not just Dyfed Regio and the eastern portions of Annwn Regio, but also the western portions of Annwn Regio, which extend across the sub-Jovian point. In fact, the absorption even appears weakly within the less-red large-scale chaos terrain near the anti-Jovian point. However, like the red color visible in imagery, this feature is absent from the chaos terrain on the leading hemisphere, which is sheltered from the trailing-hemisphere sulfur implantation and the resultant sulfur radiolytic chemistry. 

Our map may reflect the same absorber as does the incomplete \textit{Galileo} NIMS 0.7/1.2 $\micron$ ratio map published previously, which highlighted some of the same regions \citep{Carlson2005}. Like the ground-based spectra, the NIMS map was interpreted to most likely reflect sulfur chains or polymers, potentially produced as part of the radiolytic sulfur cycle on the trailing hemisphere. However, as the absorber and the reddish color with which it correlates appear so specifically associated with geologic features, we suggest that a radiolytically altered endogenous material better explains the observed geography. We discuss possible candidates in the next section.

\section{Discussion: potential compositions}\label{sec:discussion}

The HST spectra of Europa's trailing hemisphere appear to reflect both endogenous and exogenous influences on the surface composition. The implantation and subsequent radiolysis of sulfur from Io almost certainly results in the formation of sulfur allotropes, such as S$_8$ and S$_4$ \citep{Steudel1986, Carlson2002, Carlson2009}, which will affect the visible spectrum and may explain the strong near-UV downturn and 530 nm feature we observe on the trailing hemisphere. Indeed, these two species have been invoked to explain similar absorption features on Io \citep{Spencer2004, Carlson2009}. However, Europa's simultaneous global association of color with geology and dichotomy of color between the leading and trailing hemispheres seems to suggest the presence of endogenous material that has been chemically altered by the exogenous sulfur radiolysis. The geographies of the 360 nm feature and of the 700 nm slope in our spectra appear most consistent with species that are radiolytically produced from a mixture of Iogenic sulfur and endogenic material. Salts from the internal ocean, which have long been considered as likely components of Europa's surface \edit1{\citep[e.g][]{McCord1998_science, McCord1999, Dalton2007, Dalton2012, Hanley2014, Shirley2016}}, are perhaps the most obvious candidates for the endogenic starting material. Though the nature of such salts is still debated, recent work utilizing \edit1{spatially resolved} ground-based near-infrared spectra has suggested that chlorides may dominate Europa's endogenic surface salts \citep{BrownHand2013, Fischer2015, Fischer2017, Ligier2016}. Specifically, \citet{BrownHand2013} proposed a conceptual model in which these hypothesized chlorides participate in the radiolytic sulfur cycle on the trailing hemisphere and convert to sulfates when irradiated in the presence of Iogenic sulfur. In this picture, endogenic chloride-rich material would persist within geologic terrain on the leading hemisphere, where it is sheltered from the incoming sulfur plasma, but become progressively altered to a more sulfate-rich composition within those terrains subjected to the sulfur radiolysis on the trailing hemisphere. \edit1{It should be noted that this hypothesis differs from that of \citet{Ligier2016}, who also hypothesized the presence of chlorinated salts using a similar near-infrared dataset to that of \citet{BrownHand2013}, but instead interpreted their data to reflect magnesium-bearing chlorinated salts within the chaos terrain of the trailing hemisphere. However, the compositions suggested by \citet{Ligier2016} result from the linear mixture modeling of largely featureless continua, rather than from the detection of distinct, compositionally diagnostic absorption features, which is necessary to unambiguously identify surface species. Indeed, the recent HST detection of a 450 nm absorption indicative of irradiated NaCl within large-scale chaos regions on the leading hemisphere \citep{TrumboEtAl2019} represents the only unambiguous detection of chlorinated salts on Europa to date and is consistent with the conceptual view laid out by \citet{BrownHand2013}}.  Thus, sulfate salts may represent a likely candidate for the altered endogenous material implied by the visible-wavelength data \edit1{of the trailing hemisphere.} 

\begin{figure}
\figurenum{5}
\plotone{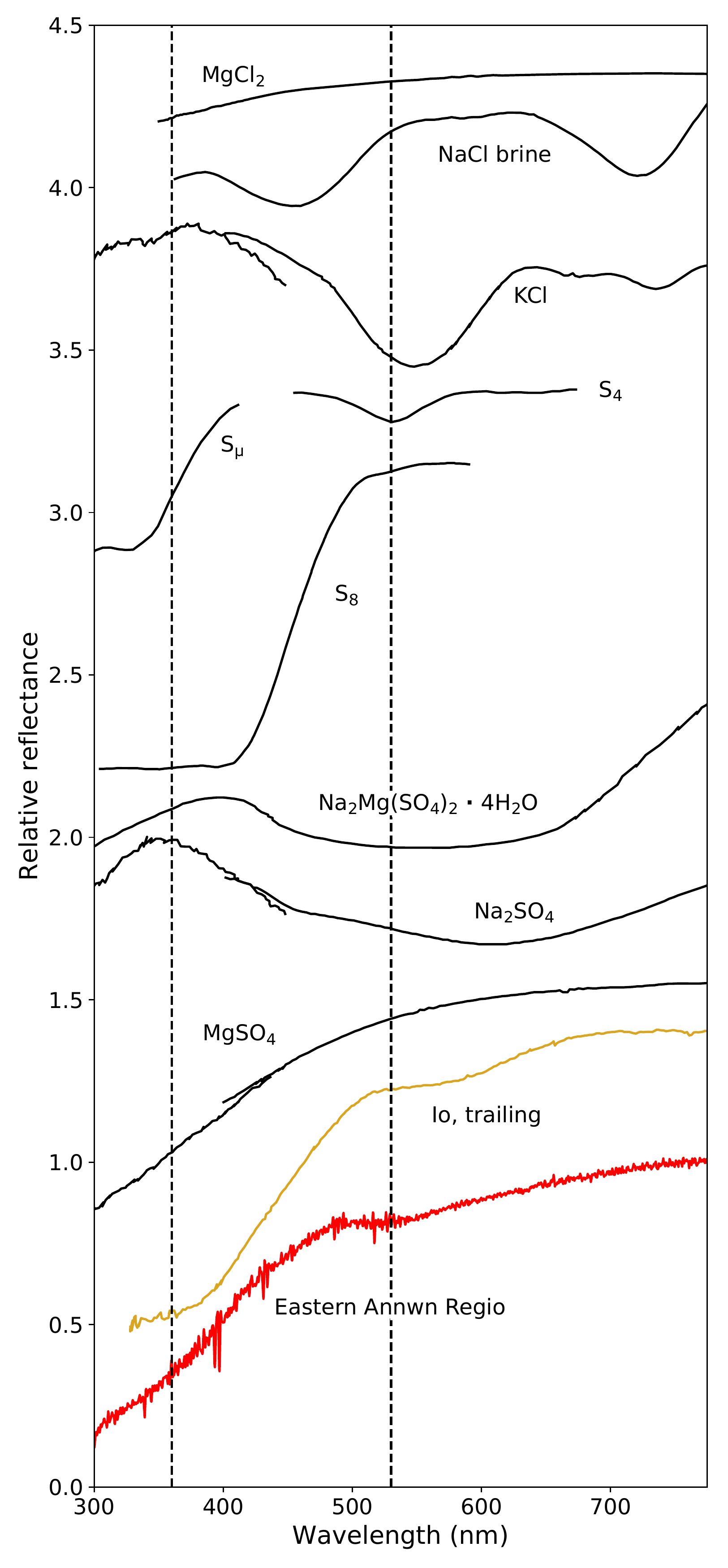}
\caption{Average spectrum of Eastern Annwn Regio compared to the spectra of multiple sulfur allotropes, select irradiated sulfate and chloride salts, and the trailing hemisphere of Io. Vertical dashed lines indicate the approximate wavelengths of the band minima for the 360 and 530 nm absorptions on Europa. With the exception of the S$_\mu$ and S$_4$ spectra, which are scaled arbitrarily for clarity, all of the spectra are scaled to unity at their longest wavelengths and offset vertically from each other. The spectra of the sulfur allotropes are adapted from \citet{Carlson2009}, the Na$_2$Mg(SO$_4$)$_2$ $\cdot$ 4H$_2$O (bloedite) spectrum is taken from \citet{NashFanale1977}, the Na$_2$SO$_4$, MgSO$_4$\edit1{, and KCl} spectra are from \citet{Hibbitts2019}\edit1{, the NaCl brine spectrum is from \citet{HandCarlson2015}, the MgCl$_2$ spectrum is taken from the supplementary materials of \citet{TrumboEtAl2019}, and the Io spectrum is from \citet{Spencer1995}. With the exception of the S$_\mu$ spectrum, all of the spectra shown represent irradiated samples. The Na$_2$Mg(SO$_4$)$_2$ $\cdot$ 4H$_2$O (bloedite) spectrum shows a proton-irradiated sample, the remaining salt spectra show electron-irradiated samples, and the S$_4$ and S$_8$ spectra are of UV-irradiated samples. With the exception of the NaCl brine spectrum, which was taken at 100 K, and the S$_8$ and S$_4$ spectra, which were obtained at 77 K, all of the shown laboratory spectra were obtained at room temperature.}\label{fig:spectra}}
\end{figure}

Though many candidate sulfate salts are typically white at visible wavelengths, like NaCl, they can become significantly discolored when subjected to radiation conditions like those at the surface of Europa \citep{NashFanale1977, Hibbitts2019}. In fact, \citet{Hibbitts2019} recently proposed that irradiated sulfate salts may explain the ground-based disk-integrated spectrophotometry of the trailing hemisphere. Specifically, \citet{Hibbitts2019} noted that irradiated MgSO$_4$, a species already suggested from the infrared spectra of \citet{BrownHand2013}, provides a decent fit to the overall shape of the trailing-hemisphere spectrum in the visible, while salts that form broad color-center absorptions near 600 nm could contribute to the apparent broad absorption beyond 500 nm, which we have shown to be a convolution of the 530 nm feature and a wider absorption spanning the red wavelengths.

Figure \ref{fig:spectra} compares an average spectrum of Eastern Annwn Regio to some of these proposed \edit1{irradiated} sulfate salts \edit1{and to select irradiated chloride salts}, as well as to the sulfur allotropes discussed in the previous section. Like the spectrum of Eastern Annwn Regio, that of irradiated MgSO$_4$ also exhibits a pronounced near-UV downturn. Thus, it is possible that MgSO$_4$ may contribute to the strong near-UV downturn we find on the trailing hemisphere, though sulfur allotropes almost certainly contribute as well and are likely required to explain the elliptic distribution we observe (Figure \ref{fig:maps}B). \edit1{Both irradiated KCl and S$_4$ exhibit absorptions nearby in wavelength to the 530 nm feature we observe on Europa. However, S$_4$ provides a more satisfactory explanation, both in terms of the wavelength of the band minimum (Figure \ref{fig:spectra}) and in terms of the geographic distribution (Figure \ref{fig:maps}D), as one would expect KCl to be spatially associated with the previously observed NaCl on the leading hemisphere \citep{TrumboEtAl2019}. Though sulfur allotropes may be implicated for the near-UV downturn and perhaps the 530 nm absorption}, color-center absorptions by \edit1{irradiated} sulfate salts similar to the shown Na$_2$SO$_4$ or Na$_2$Mg(SO$_4$)$_2$ $\cdot$ 4H$_2$O (bloedite) may better explain the broad absorption causing the observed spectral slope at 700 nm (Figure \ref{fig:maps_color}), which maps to the reddish material visible in imagery. However, these laboratory spectra bear little resemblance to the Europa spectrum beyond both exhibiting broad features across the red wavelengths. Thus, a conclusive correspondence between sulfate color centers and the Europa spectra is by no means implied from the available data. In fact, it is impossible to either identify or rule out any of the sulfates shown, due to the broad nature of their absorption features, the interference of multiple features within the Europa spectra, and the limitations of the laboratory data, which were obtained at room temperature using unrealistically high radiation fluxes. Furthermore, though our observed geography of the 360 nm feature on Europa suggests that it too results from altered endogenous material, none of the examined laboratory spectra provide a satisfactory explanation for this absorption. Thus, while we, in part, agree with \citet{Hibbitts2019} and suggest that irradiated sulfate salts may explain those aspects of the visible Europa spectra that correlate with geologic features on the trailing hemisphere, a better understanding of the surface composition and sulfur radiolysis chemistry and additional laboratory spectra are needed to fully address this hypothesis. 

\section{Conclusions}\label{sec:conclusions}

Utilizing spatially resolved visible-wavelength spectra of Europa from HST, we have examined several absorption features unique to the trailing hemisphere in an attempt to disentangle potential endogenous influences from those of the exogenous radiolytic sulfur chemistry. By comparing the distribution of each absorption with surface color, geology, and radiation bombardment patterns, we differentiate between features that we interpret to reflect pure-sulfur radiolytic products and those that we interpret to reflect species radiolytically produced from a combination of endogenic material and Iogenic sulfur. Two of the features we observe---a widespread near-UV downturn and a distinct feature at 530 nm---appear consistent with sulfur allotropes, as has been suggested based on previous ground-based data. However, the geographies of the remaining features---a discrete absorption at 360 nm and the spectral slope at red wavelengths---appear to indicate endogenous material altered by sulfur radiolysis. Though we cannot uniquely identify the responsible species with currently available data, we suggest irradiated sulfates produced by the radiolysis of endogenous salts as potential candidates. We suggest that future laboratory experiments examining the sulfur radiolysis of potentially endogenous salts and investigating the spectroscopy of irradiated sulfates at Europa-like temperatures and energy fluxes may provide further insight in to the interpretation of the HST spectra.

\acknowledgements
This work is based on observations made with the NASA/ESA HST, obtained at the Space Telescope Science Institute, which is operated by the Association of Universities for Research in Astronomy Inc., under NASA contract NAS 5-26555. These observations are associated with program number 14650. This work was supported by NASA Headquarters under the NASA Earth and Space Science Fellowship Program (grant 80NSSC17K0478). Support for this work was provided by NASA through grant number HST-GO-14650.001-A from the Space Telescope Science Institute, which is operated by AURA Inc., under NASA contract NAS 5-26555. This research was supported by grant 1313461 from the NSF. K.P.H. acknowledges support from the Jet Propulsion Laboratory, California Institute of Technology, under a contract with NASA and funded, in part, through the internal Research and Technology Development program.

\end{document}